\documentclass[runningheads]{llncs}
\usepackage{graphicx}
\usepackage{hyperref}
\usepackage{soul}
\usepackage{color}
\usepackage{xcolor}
\graphicspath{ {./images/} }
\begin{document}
%

\title{Towards peer-to-peer  sharing of wireless energy services}

%
%
\author{Pengwei Yang\and Amani Abusafia \and
Abdallah Lakhdari\and 
Athman Bouguettaya}
\authorrunning{P. Yang et al.}
\institute{The University of Sydney, Sydney NSW 2000, Australia\\
\email{pyan8871@uni.sydney.edu.au,\\\{amani.abusafia,abdallah.lakhdari,athman.bouguettaya\}@sydney.edu.au}}

\maketitle              
\vspace*{-20pt}
\begin{abstract}

Crowdsourcing wireless energy services is a novel convenient alternative to charge IoT devices. We demonstrate\textit{ peer-to-peer wireless energy services sharing between smartphones} over a \textit{distance}. Our demo leverages  (1) a service-based technique to share energy services, (2)  state-of-the-art power transfer technology over a distance, and (3) a mobile application to enable communication between energy providers and consumers. In addition, our application monitors the charging process between IoT devices to collect a dataset for further analysis. Moreover, in this demo, we compare \textit{the peer-to-peer energy transfer between two smartphones using different charging technologies}, i.e., cable charging, reverse charging, and wireless charging over a distance.  A set of preliminary experiments have been conducted on a real collected dataset to analyze and demonstrate the behavior of the current wireless and traditional charging technologies.


\end{abstract}
\vspace{-30pt}
\section{Introduction}
\vspace{-10pt}

Wireless power transfer (WPT), also known as wireless charging, has been widely adopted as a flexible and ubiquitous solution to charge IoT devices \cite{fang2018fair} \cite{sakai2021towards}. Several studies leveraged  the service paradigm to propose the concept of crowdsourcing wireless energy as \textit{energy services} to charge nearby IoT devices \cite{abusafia2022services} \cite{lakhdari2021fairness} \cite{abusafia2022quality}. Energy Service is defined as the abstraction of the wireless energy transfer from one IoT device (i.e., \textit{provider}) to another device (i.e., \textit{consumer}) \cite{lakhdari2021proactive}. Energy services may be shared wirelessly with the development of new technologies known as “Over-the-Air” wireless charging \cite{sakai2021towards} \cite{lakhdari2021fairness}.

Recent studies have made substantial progress in addressing several challenges in crowdsourcing energy services, including service composition, fairness and quality of experience \cite{lakhdari2021fairness} \cite{abusafia2022services} \cite{abusafia2022quality} \cite{abusafia2022maximizing}. However, existing works rely on simulation analysis of the proposed techniques using synthetic datasets. In other words, there is no demonstration of the existing literature on \textit{real devices using wireless power transfer over a distance.} Hence, in this paper, we demonstrate peer-to-peer sharing of wireless energy services among smartphones over a distance. Our demo leverages a service-based technique to share energy services, the state-of-the-art power transfer technology over a distance, and an extension of the proposed app by \cite{YaoJessica2022WIES}. Similar to the previously mentioned app, our app enables consumers and providers to share energy conveniently and to monitor the sharing process for further analysis. In addition, unlike the previous app, our demo allows users to (1) share energy by requesting an amount of energy or requesting to be charged for a certain time period and (2) transfer energy wirelessly over a distance. Moreover, in this demo, we conduct preliminary experiments to verify the feasibility and stability of the platform with wireless charging over a distance. We also compare and analyze the performance of the wireless charging technology with reverse and cable charging technologies.

\begin{figure}[!t]
    \centering
       \setlength{\abovecaptionskip}{-3pt}
    \setlength{\belowcaptionskip}{-20pt} \includegraphics[width=0.55\linewidth]{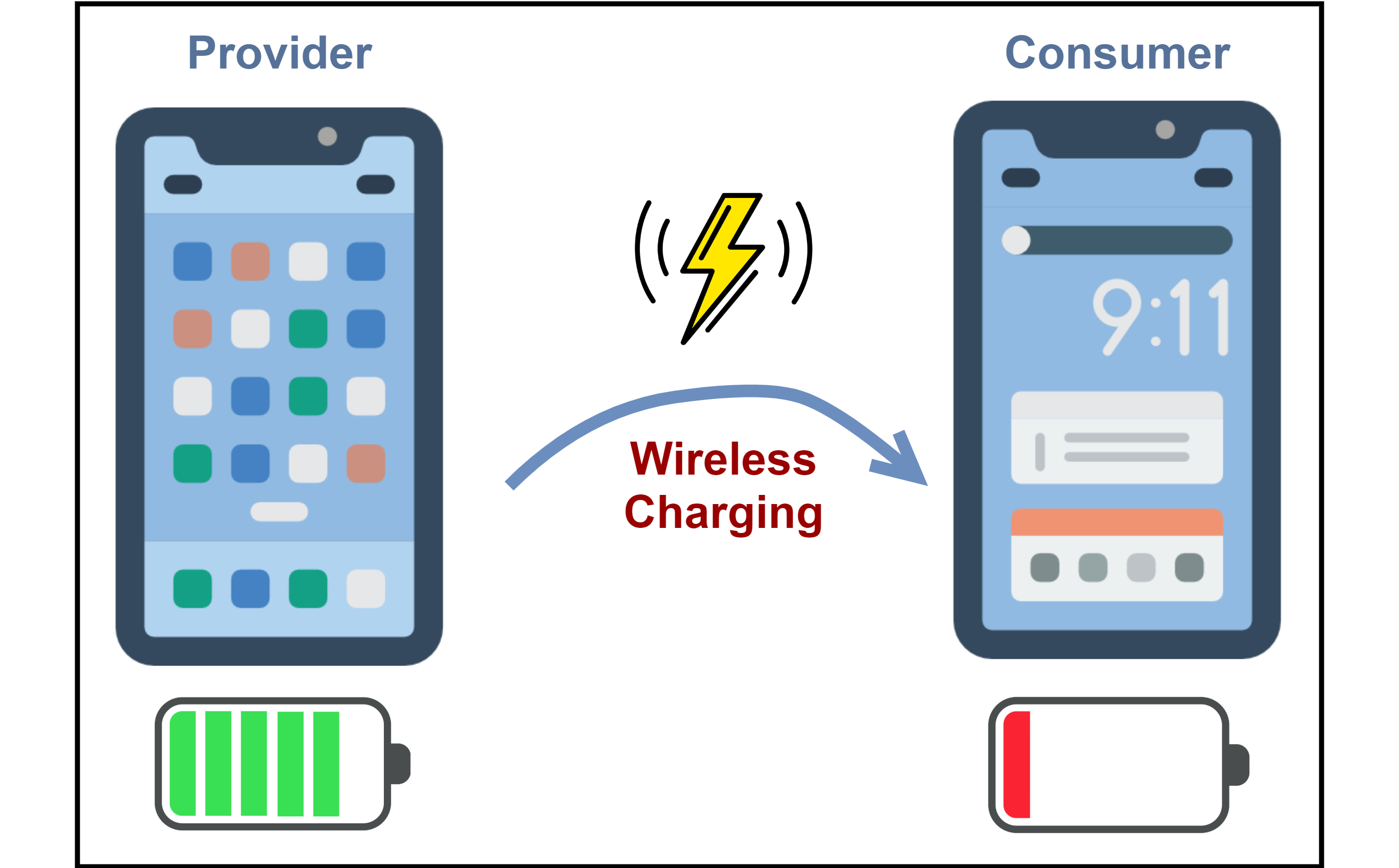}    
    \caption{Wireless energy charging scenario}
    \label{synthesis}

\end{figure}


\vspace{-15pt}
\section{System Overview}
\vspace{-10pt}
We consider a scenario where people congregate in confined areas (e.g., restaurants, coffee shops, and movie theaters). In the confined areas, we assume that IoT users will share their spare energy or request energy with nearby devices (See Fig.\ref{synthesis}). IoT users will use our mobile application to request energy as \textit{energy consumers} or to accept energy requests as \textit{energy providers}. In our app, IoT users send and receive requests directly using Bluetooth. Moreover, our mobile application extends the app proposed by \cite{YaoJessica2022WIES}  because of its limitations in terms of synchronization, variety of request types, and using built-in reverse charging. In contrast, our application expends the previous application with the following features: (1) our app allows users to send and receive energy based on size. For instance, a consumer may ask a provider to charge 1000 mAh. (2) our app also offers the option of requesting energy based on a time period. For instance, a consumer may ask a provider to charge them for the next 10 minutes. Finally, (3) our application records the battery status of providers and consumers every time interval $t$, e.g., every 10 seconds, with the actual time of each record. This enables us to synchronize the monitoring process by ensuring that the data is recorded at the exact time, e.g., second. The recorded data may be used for further analysis, e.g., studying the charging rate or optimizing the energy loss.

\looseness=-1 
\vspace{-15pt}
\section{Demo Scenario}
\label{sc}
\vspace{-10pt}
Our application runs in the following scenario. Once a consumer has submitted a request. Our platform will automatically submit a request to the closest provider and wait for the provider's response. If the provider accepts,  the energy sharing and monitoring process will start. Then the process will end once the consumer receives their request (e.g.,  an amount of energy or being charged for a specific period). Simultaneously, the platform synchronizes the monitoring process between the provider and consumer and captures the data of the entire energy transfer process. Once the sharing process ends, the app uploads the collected data to the edge service management system (e.g., a router associated with the confined area). The current version of our energy-sharing platform supports a one-to-one energy transfer mode and has demonstrated reliable performance in monitoring and demonstrating the wireless energy-sharing process over a distance.\looseness=-1 

\begin{figure}[!t]
    \centering
    \setlength{\abovecaptionskip}{-5pt}
\setlength{\belowcaptionskip}{-10pt}
    \includegraphics[width=1\linewidth]{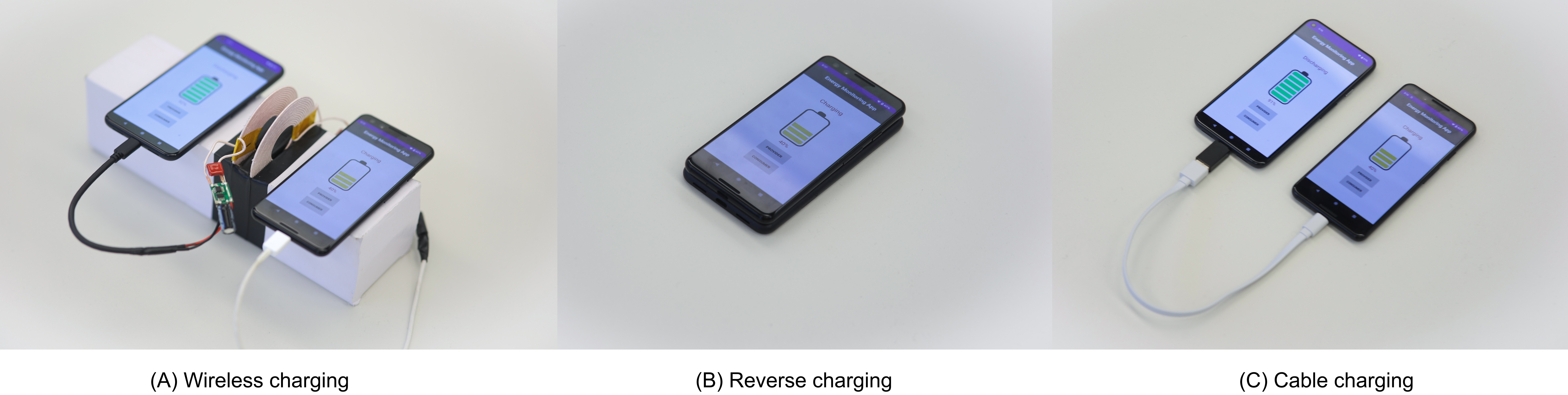}
    \caption{Used charging technologies}
    \label{device}

\end{figure}
\vspace{-20pt}
\section{Demo Setup}
\label{demo}
\vspace{-10pt}

Our demo will exhibit our platform through a real case of wireless charging over a distance between two smartphones. In this case, we use near-field wireless charging (i.e., inductive coupling wireless technique). The inductive coupling device consists of two coils, and the electrical energy is delivered based on the magnetic field (See Fig.\ref{device} (A)). We use a Google Pixel 5 smartphone as the provider and a Google Pixel 3 as the consumer. We connect both the consumer and provider to the wireless charging coils, and then we run a case similar to the previously mentioned scenario in Sec.\ref{sc}. Additionally, we will display a video of the entire process of using the devices and app to request, receive and monitor the energy-sharing process in real time. The video is published at this link: youtu.be/d-bdFGk6z4A. We will also bring other charging technologies for the visitors to our booth to try with our app (See Fig.\ref{device} (B) and (C)).

\begin{figure}[!t]
    \centering
      \setlength{\abovecaptionskip}{-3pt}
    \setlength{\belowcaptionskip}{-20pt} 
    \includegraphics[width=0.9\linewidth]{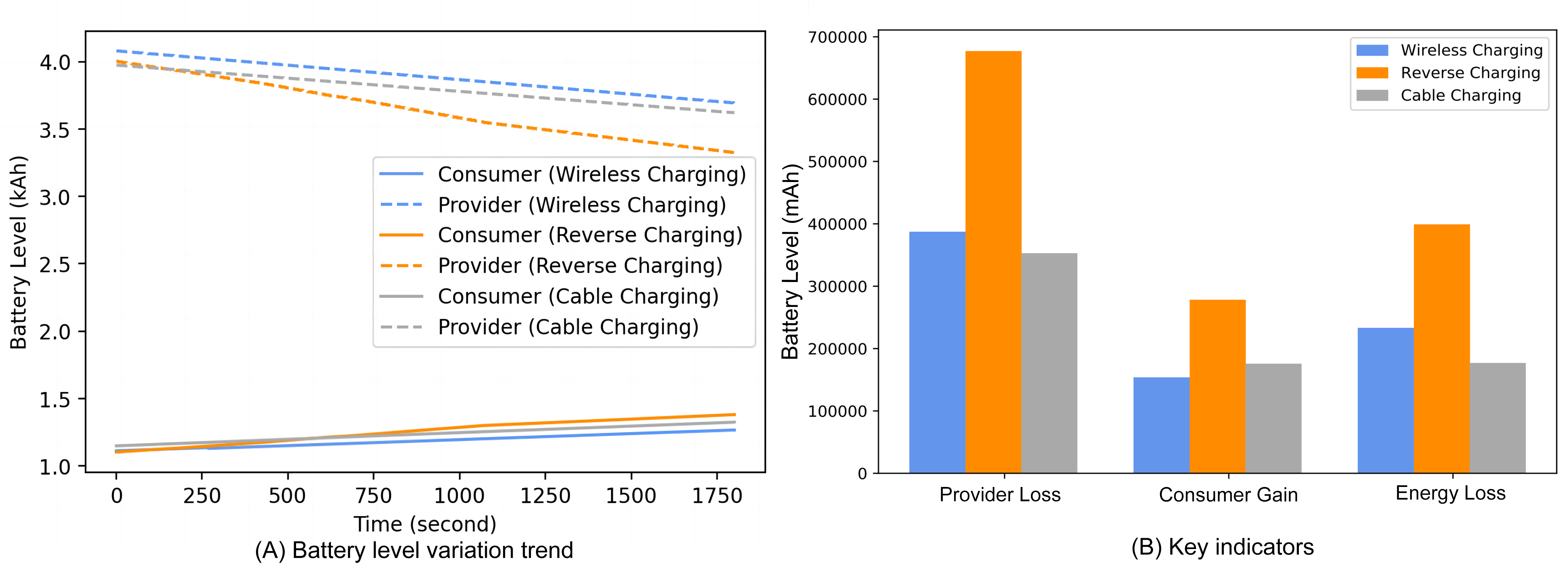}
    \caption{Comparison among different charging technologies}
    \label{data}
\end{figure}

\vspace{-20pt}
\section{Preliminary Experiments}
\vspace{-10pt}
We conducted preliminary experiments to demonstrate our platform's feasibility and monitor different charging processes. We used the same setup mentioned in Sec.\ref{demo}. We ran two experiments on the collected dataset from the charging process to analyze and demonstrate the behavior of each charging technology, i.e., wireless charging, reverse charging, and cable charging (See Fig.\ref{device}). In wireless charging, the distance between the two coils is set to 2 centimeters. The first set of experiments aims to compare differences between charging technologies in terms of battery level variation trend, provider loss, consumer gain, and energy loss. The start battery of the consumer was 40\%; the provider battery was 100\%. We ran the experiment using a time-based request to receive energy. The requested time was 30 minutes. Additionally, we set the recording interval to 1 second. The experiment shows that all approaches have similar behavior (See Fig.\ref{data}). However, the provider with reverse charging had the highest energy loss. It is worth noting that energy loss involves the energy consumed to share energy and to run the device, e.g., run the operating system, system applications, etc.

\begin{figure}[!t]
    \centering
       \setlength{\abovecaptionskip}{-3pt}
    \setlength{\belowcaptionskip}{-20pt}
    \includegraphics[width=0.9\linewidth]{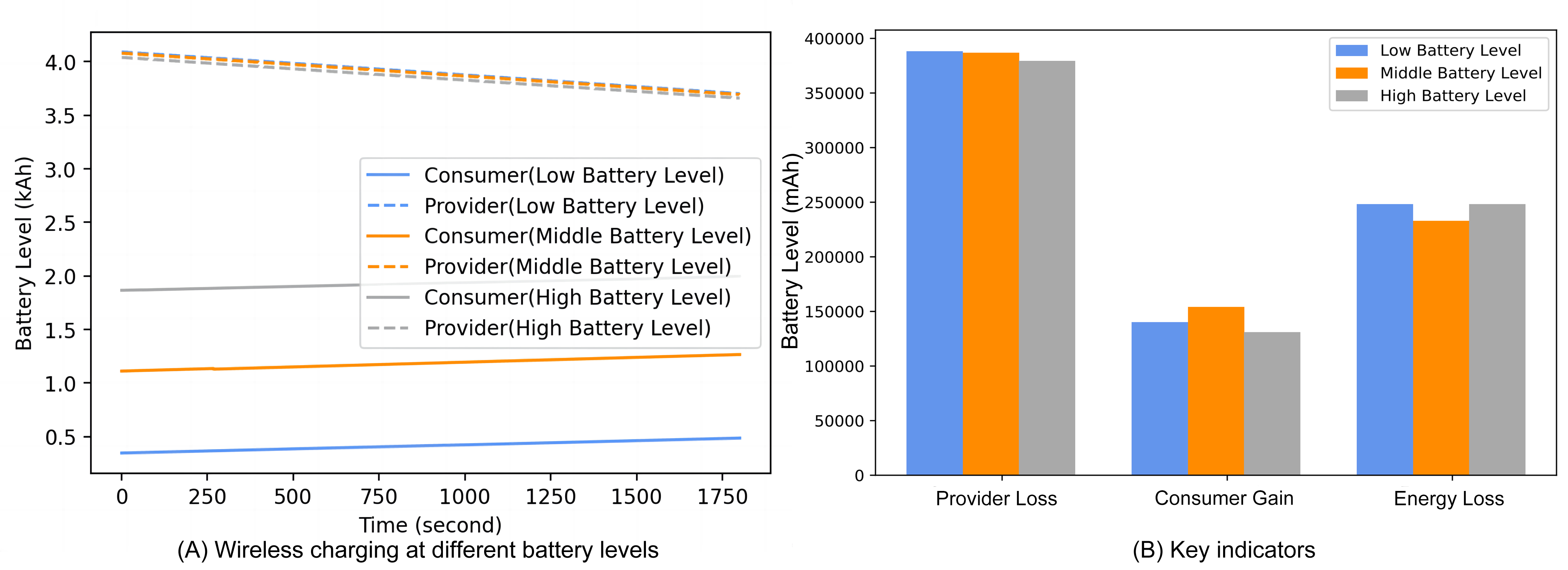}
    \caption{Comparison of wireless charging with different battery levels for consumer}
    \label{data2}
\end{figure}
The second set of experiments aims to test wireless charging with different battery levels for the consumer at each experiment. The consumer's start battery level was tested on 10\%, 40\%, and 70\% and labeled as low, middle, and high battery levels. The experiment shows that varying battery levels have less impact on the charging process (See Fig.\ref{data2}).
\vspace{-20pt}
\section*{Acknowledgment}
\vspace{-10pt}
This research was partly made possible by  LE220100078 and LE180100158 grants from the Australian Research Council. The statements made herein are solely the responsibility of the authors.
\vspace{-10pt}

\bibliographystyle{splncs04}
\bibliography{main.bib}

\end{document}